\documentstyle[aps,epsfig]{revtex}
\input epsf
\def\be{\begin{equation}}
\def\ee{\end{equation}}
\begin{document}
 \title{Density-Matrix Spectra of Solvable Fermionic Systems}
 \author{
 Ming-Chiang Chung
  and Ingo Peschel \\
{\small Fachbereich Physik, Freie Universit\"at Berlin,} \\
{\small Arnimallee 14, D-14195 Berlin, Germany}}
 \maketitle
 \begin{abstract}
 We consider non-interacting fermions on a lattice and give a general result
 for the reduced density matrices corresponding to parts of the system. This
 allows to calculate their spectra, which are essential in the DMRG method,
 by diagonalizing small matrices. We discuss these spectra and their typical
 features for various fermionic quantum chains and for the two-dimensional
 tight-binding model.
\end{abstract}
  
 \section{Introduction}
 
Density matrices have found an interesting new application in recent years.
In the density-matrix renormalization group (DMRG) method 
\cite{White92,White93,DMRG} 
a quantum system is built from smaller parts and the idea is to work with
basis functions in these parts which are optimal for the combined system.
These are the eigenfunctions of the reduced density matrices which have the
largest eigenvalues $w_n$. Obviously, the procedure will work well if the
eigenvalue spectrum drops rapidly, such that a small number of functions
practically exhausts the sum rule ${\small \sum} w_n =1$. For quantum chains this is 
indeed the case. The numerical calculations show a roughly exponential 
decrease of the eigenvalues \cite{White93,Kaulke98}. Of course, this raises
the question whether such spectra can be obtained explicitely
for some solvable models. For non-critical systems this is possible by using
the relation \cite{Nishino96} between the density matrices of quantum 
chains and the corner transfer matrices (CTM's) \cite{Baxter} of the corresponding 
two-dimensional classical problems.
 In this way, the spectra for the transverse 
Ising chain \cite{Pescheletal99}, 
the XXZ Heisenberg chain \cite{Pescheletal99} 
and a chain of coupled oscillators \cite{Peschel/Chung99}
 could be determined in
the thermodynamic limit and compared with DMRG calculations. 
In all these cases, 
one finds simple analytic expressions and, apart from degeneracies, a strictly 
exponential behaviour. This does not hold for the chiral 
three-state Potts chain 
\cite{Ritter/Gehlen00} or for non-integrable 
models \cite{Ritter99,Okunishi99}, 
but qualitatively the spectra are similar.

Given the importance of fermionic systems in general 
and also for DMRG applications,
one would of course like to have results for this case, too. 
The transverse Ising chain can 
be viewed as a fermionic model, but the CTM approach does not make use of this 
and is limited to large non-critical systems.  
Therefore an alternative approach
is necessary by which one can treat solvable fermion systems of arbitrary size.
In the present communication we show how this can be done. 
The systems which we consider are non-interacting, such that the Hamiltonian
can be diagonalized by a Bogoliubov transformation. Using an explicit form
of the state in question (usually the ground state), we show
that arbitrary reduced density matrices can be calculated exactly and have the
general form $exp(-{\cal H})$. The operator ${\cal H}$
describes a collection of new,
non-interacting fermions with single-particle eigenvalues $\varepsilon_l$. 
Apart from the different statistics, this is the same situation as for coupled 
oscillators \cite{Peschel/Chung99,Chung/Peschel00}. 
The $\varepsilon_l$, which determine the properties of the
spectrum, follow from the eigenvalues of an $M \times M$ matrix, 
where $M$ is the
number of sites in the chosen subsystem. 
In general, they have to be calculated 
numerically. One should stress that the dimensionality of the system plays no
essential role.

In the following Section \ref{sec:two} we sketch 
the general method of computation which uses 
coherent fermionic states for calculating the necessary partial traces. 
In Section \ref{sec:three},
we apply it to the transverse Ising chain and discuss the resulting
spectra for a number of situations, including the critical
case, the first excited state and
related row transfer matrices. Section \ref{sec:four} deals 
briefly with another one-dimensional problem, namely the
spin one-half XY-chain in a field. This is interesting because it has
a disorder point where the spectrum collapses. 
In Section \ref{sec:five} we turn to the
physically most important case of a tight-binding model which we dicuss in
two dimensions. We present spectra for systems of various sizes and shapes, as
well as truncation errors showing the difficulties in this case.  
Section \ref{sec:six}, 
finally, contains a summary and some additional remarks. 
Some technical details 
can be found in the Appendix.

 \section{Method}\label{sec:two}

 We consider Hamiltonians which are quadratic
  in Fermi operators and thus have the general form 
   \begin{equation} 
    H \;=\; \sum_{ij=1}^{L} [ c_i^{\dagger} A_{ij} c_j+
         \frac{1}{2} (c_i^{\dagger} B_{ij} c_j^{\dagger} + h.c.) ]
       \label{eqn:Ham}
    \end{equation}  
  where the $c_i$'s and $c_i^{\dagger}$'s are Fermi annihilation and 
  creation operators. Because of the Hermiticity of $H$, the matrix
  {\boldmath$A$} is Hermitian and {\boldmath$B$} is antisymmetric.
  In the following we consider only real matrices. 
  One can 
  diagonalize $H$ through the canonical transformation\cite{Lieb61} 
  \begin{equation} 
   \eta_k\;=\;\sum_i (g_{ki} c_i + h_{ki} c_i^{\dagger})
  \label{eqn:bogoliubov}
  \end{equation}
  which leads to
  \begin{equation}
   H\;=\;\sum_k \Lambda_k \eta_k^{\dagger} \eta_k + \mbox{constant}
  \end{equation}
  The quantities  $\Lambda_k^2$ are the eigenvalues of the matrices 
  ({\boldmath$A-B$})({\boldmath$A+B$})
  and  {(\boldmath$A+B$)}{(\boldmath$A-B$)},
  the corresponding eigenvectors being $\phi_{ki} = g_{ki}+h_{ki}$
  and $\psi_{ki} = g_{ki}-h_{ki}$,respectively.

  Consider now the ground state $\mid\Phi_0>$ of the Hamiltonian (\ref{eqn:Ham})
  for an even number of sites $L$. Due to the structure of $H$, it is
  a superposition of configurations with either an even or an odd
  number of fermions. This suggests to write it (for the even case)
  in the form  
  \begin{equation} 
   \mid\Phi_0> \;=\;C \;\exp{\{\frac{1}{2}
               \sum_{ij} G_{ij} c_i^{\dagger} c_j^{\dagger}\}} \mid 0>
   \label{eqn:gsconf}
  \end{equation}       
  where $\mid 0>$ is the vacuum of the $c_i$, i.e.
  \begin{equation}
   c_i\mid 0> = 0
  \end{equation}
  Such an exponential form is known from superconductivity, where
  the BCS wave function (in momentum space) can be written in this way
  \cite{Parks69}.

  One obtains $G_{ij}$ by 
  applying the Fermi operators $\eta_k$ 
  to the ground state 
  \begin{equation}
   \eta_k \mid\Phi_0> \;=\;0 \;\;\;\;\; \mbox{for all $k$}
   \label{eqn:eta}
  \end{equation}
  which leads to (see Appendix)
  \begin{equation} 
   \sum_{m} g_{km}G_{mn} + h_{kn} = 0\;\;\;\;\; \mbox{for all $k,n$}
  \label{eqn:gsrelation}
  \end{equation}
  Thus {\boldmath$G$} relates the two matrices {\boldmath$g$} and 
  {\boldmath$h$} 
   of the transformation (\ref{eqn:bogoliubov}).
  Using (\ref{eqn:gsconf}), one obtains the total density matrix
   $\rho_0 = \mid\Phi_0><\Phi_0\mid$
  explicitly in an exponential form
  \begin{equation}
   \rho_0= {|C|}^2 \exp{(\frac{1}{2} \sum_{ij} G_{ij} 
   c_i^{\dagger} c_j^{\dagger})}
   \mid 0><0\mid\exp{(-\frac{1}{2} \sum_{ij} G_{ij} c_i c_j)}  
  \label{eqn:rho0}
  \end{equation}

  One now divides the total system in two parts
  (system and environment in the DMRG terminology) and 
  looks for the reduced density matrix  in part 1. 
  This is obtained by taking the trace over part 2.
  \begin{equation}
    \rho_1=Tr_2 \;(\rho_0)
  \end{equation}  
  In order to calculate $\rho_1$, one uses the fermionic coherent states 
  defined by \cite{Negele87}
  \begin{equation}
   c_i\mid\xi_1\cdots\xi_L>=\xi_i\mid\xi_1\cdots\xi_L> 
  \end{equation}
   Such states can be built from the vacuum
  with operators $c_i$ and Grassmann variables $\xi_i$
  \begin{equation} 
   \mid\xi_1\cdots\xi_L>=\exp{(-\sum_i \xi_i c_i^{\dagger})}\mid 0>
    \label{eqn:coherent}
  \end{equation}
  Using this, one can write the trace of an operator $O$ as
  \begin{equation}
   Tr\;O = \int \prod_{\alpha}d\xi_{\alpha}^{\ast}d\xi_{\alpha}
    e^{-\sum_{\alpha} \xi_{\alpha}^{\ast}\xi_{\alpha}} <-\xi\mid O \mid\xi>
   \label{eqn:trace}
  \end{equation}
  
  After forming a general matrix element of $\rho_0$ with such states
  and  taking the trace over the environment  with (\ref{eqn:trace}),
  one obtains, if part 1 consists of $M$ sites  
  \begin{eqnarray}
      &&<\xi_{1}\cdots\xi_{M}\mid\rho_1
    \mid\xi'_1\cdots\xi'_{M}>\nonumber \\
   && = {|C|}^2 \int\prod_{i=M+1}^{L} 
      d\xi_i^{\ast}d\xi_i
     e^{-\sum_{i}\xi_i^{\ast}\xi_i}
     <\xi_1\cdots\xi_{M}-\xi_{M+1}
      \cdots-\xi_{L}\mid\rho_0\mid\xi'_1\cdots
     \xi'_{M}\xi_{M+1}
      \cdots\xi_{L}> \label{eqn:integral}
    \end{eqnarray} 
    Inserting (\ref{eqn:rho0}) leads to  an integrand  
    which contains only quadratic forms of Grassmann variables 
    in the exponents. The integration can then be carried out   
    by rotating and displacing 
    the variables as for a Gaussian integral with  complex numbers. 
    This gives
  \begin{eqnarray}
   &&<\xi_{1}\cdots\xi_{M}\mid\rho_1
    \mid\xi'_1\cdots\xi'_{M}> \nonumber\\
   && = |C|^2 \exp{(\sum_{ij} \alpha_{ij} \xi_i^{\ast}\xi_j^{\ast})}  
      \exp{(\sum_{ij} \beta_{ij} \xi_i^{\ast}\xi'_j)}
    \exp{(\sum_{ij} -\alpha_{ij} \xi'_i\xi'_j)} \hspace{1cm};\hspace{0.5cm} 
     i,j\leq M \label{eqn:gve}   
  \end{eqnarray}
  The $M\times M$ matrices $\alpha$ and $\beta$ appearing here
  are defined as follows. One divides {\boldmath$G$} into four submatrices
  $a^{11},a^{12},a^{21}$ and $a^{22}$, according to whether 
  the sites $i,j$ belong to part 1 or part 2. In terms of these
  \begin{eqnarray}
   2\; \alpha & = & a^{11} + c a^{22} c^{T} \nonumber\\
   \ \; \beta & = & c c^{T}
  \end{eqnarray}  
  where $c=a^{12}(1-a^{22})^{-1}$ and $c^{T}$ denotes its transpose. 
  As shown in the Appendix one can reconstruct the operator form of
   $\rho_1$ from 
  the matrix elements (\ref{eqn:gve}) . This gives 
   \begin{equation} 
    \rho_1= |C|^2 \exp{(\sum_{ij} \alpha_{ij} c_i^{\dagger}c_j^{\dagger})}  
      \exp{(\sum_{ij} (\ln{\beta})_{ij} c_i^{\dagger}c_j)}
    \exp{(\sum_{ij} -\alpha_{ij} c_i c_j)} \hspace{1cm};\hspace{0.5cm} 
     i,j\leq M    
    \label{eqn:rho1o}
    \end{equation}
  Finally, since the Fermi operators appear quadratic  in the exponents,
  $\rho_1$ can be diagonalized with a Bogoliubov transformation 
  as in (\ref{eqn:bogoliubov}). As a result,
   \begin{equation}
   \rho_1= K\; \exp{(-\sum_{l=1}^{M} \varepsilon_l f_l^{\dagger} f_l)}
   \label{eqn:rho1}
  \end{equation} 
  with new Fermi operators $f_l^{\dagger},f_l$ and $K=|C|^2$.
  The single-particle eigenvalues
  $\varepsilon_l$ follow from the matrices $\alpha, \beta$
  according to Eqn. (\ref{eqn:eigen}) of the Appendix.
  The normalization factor $K$ is fixed by the sum rule 
  $Tr (\rho_1) =1$. 
   In this way, one can calculate the  density-matrix
  spectra numerically for an arbitary part of a finite 
  system with Hamiltonian (\ref{eqn:Ham}).

\begin{figure}
 \hspace*{4cm}
     \epsfxsize=70mm
     \epsfysize=60mm
     \epsffile{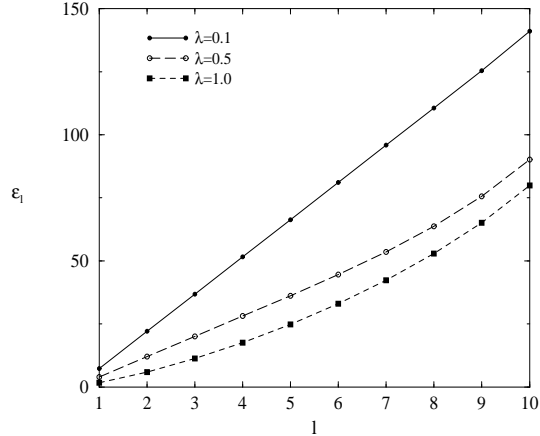}
    \caption{Single-particle eigenvalues $\varepsilon_l$ 
     for one half of a transverse Ising chain, arranged in ascending order.
     The system is in the groundstate, $L=20$ and $\lambda < 1$.}
     \label{fig1}
\end{figure}

\begin{figure}
 \hspace*{4cm}
     \epsfxsize=70mm
     \epsfysize=60mm
     \epsffile{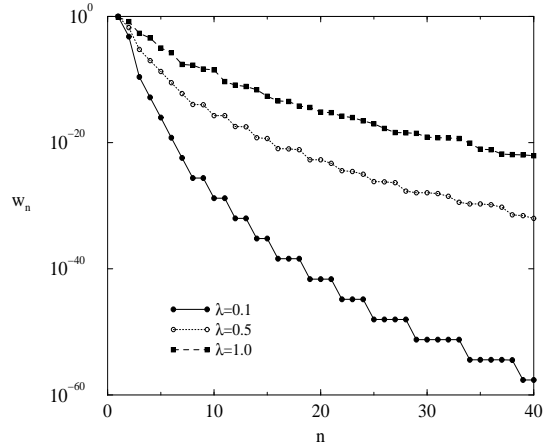}
\caption{Density-matrix eigenvalues $w_n$, arranged in decreasing order,
  obtained from the $\varepsilon_l$
  in Fig. \ref{fig1} and for the same parameters.
 } \label{fig2}
\end{figure}
         
\section{Transverse Ising Chain}
  \label{sec:three}
  As a first example,we consider 
  in this section the transverse Ising chain 
  with open boundaries described by
  \be
   H = -\sum_{i=1}^{L}\sigma^z_i-\lambda \sum_{i=1}^{L-1}
   \sigma^x_{i} \sigma^x_{i+1} 
   \label{eqn:transI}
  \ee 
   where the $\sigma^{\alpha}$ are  Pauli spin matrices and
   the transverse field has been set equal to one. In the thermodynamic limit,
   this system has a quantum critical point at $\lambda=1$ and long-range
   order in $\sigma^x$ for $\lambda > 1$. In terms of spinless fermions H
   reads
   \be
    H = -2\sum_{i=1}^{L} (c_i^{\dagger}c_i-1/2)-
   \lambda\sum_{i=1}^{L-1} (c_i^{\dagger}-c_i)(c_{i+1}^{\dagger}+c_{i+1})
   \label{eqn:fTIM}
   \ee
   and thus has the form (\ref{eqn:Ham}). In the following we discuss
   the reduced density matrix $\rho_1$ for one half of the chain, i.e. 
   $M = L/2$.
   
  We first consider the ground state.
  In figure \ref{fig1}, 
  the single-particle eigenvalues $\varepsilon_l$ are plotted
  for $L=20$ and different coupling constants $\lambda$. For $\lambda=0.1$ 
  they all lie on a straight line, which corresponds to the situation one finds
  in the thermodynamic limit. This is what one expects 
  since the correlation length
  is much less than $L$ and hence boundary effects should be small. 
  One can also
  check that the values are exactly those obtained analytically via corner 
  transfer matrices \cite{Pescheletal99}. It seems to be difficult, however,
  to derive these results directly from our equations. For larger coupling, 
  $\lambda = 0.5$, only the first
  $\varepsilon_l$ follow a linear law, then the curve bends upwards.
  This is similar to the behaviour one finds for finite-size corner transfer
  matrices \cite{Truong87}, although the geometry there is 
  different. At the same time, the initial slope decreases. Finally, at 
  the critical point, the whole graph is curved. In the ordered region
  (not shown), a linear regime develops again.

  From the $\varepsilon_l$ one obtains the actual
  eigenvalues $w_n$ of $\rho_1$ by specifying the occupation numbers 
  $f_l^{\dagger}f_l$ in (\ref{eqn:rho1}). The resulting spectra are shown in 
  figure \ref{fig2} in a semi-logarithmic plot. Note that not all $w_n$ are
  shown, however they are correctly normalized to one. Similar results, but
  for a smaller number of $w_n$, were obtained in \cite{Pescheletal99} 
  via DMRG calculations. Due to the relatively
  large values of the $\varepsilon_l$ there is a rather rapid
  decay (note the vertical scale) so that the system can be treated
  very well by DMRG \cite{Legeza96,Sznajd00}.
  This holds even at the critical point, where
  the decay is slowest.
   
  The situation there is presented in more detail in the next figures.
  Figure \ref{fig3} shows the
  $\varepsilon$-spectra for various sizes of the system.
  As $L$ increases, the number of $\varepsilon$ increases, the
  curves become flatter, but the curvature remains. There is no sign
  of a linear region related to conformal invariance on this scale
  (compare \cite{Truong87}). 
  The $w_n$ spectra are plotted in figure \ref{fig4}.
  Because of the form of the $\varepsilon$, there are few degeneracies
  and the curves have the typical, relatively smooth shape found also
  for other critical systems \cite{White93,Kaulke98}. The finite-size
  effects show up essentially in the tails.

\begin{figure}
 \hspace*{4cm}
     \epsfxsize=70mm
     \epsfysize=60mm
     \epsffile{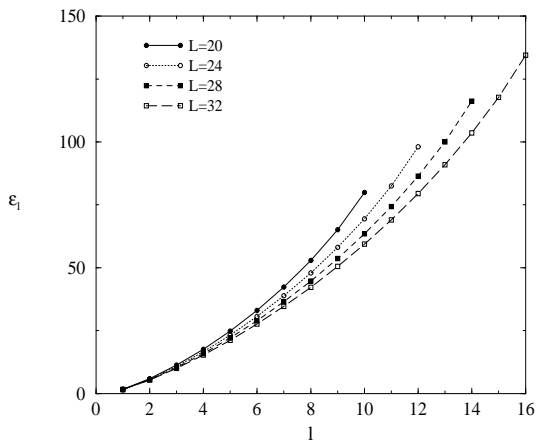}
\caption{Single-particle eigenvalues $\varepsilon_l$ 
 for critical transverse Ising chains in the ground state.} \label{fig3}
\end{figure}

\begin{figure}
 \hspace*{4cm}
     \epsfxsize=70mm
     \epsfysize=60mm
     \epsffile{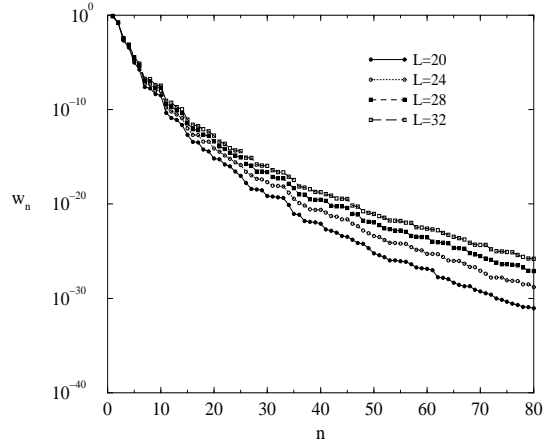}
\caption{Density-matrix eigenvalues $w_n$ for transverse Ising chains 
 at the critical point obtained
  from the $\varepsilon_l$ in Fig. \ref{fig3}.} \label{fig4}
\end{figure}

\begin{figure}
 \hspace*{4cm}
     \epsfxsize=70mm
     \epsfysize=60mm
     \epsffile{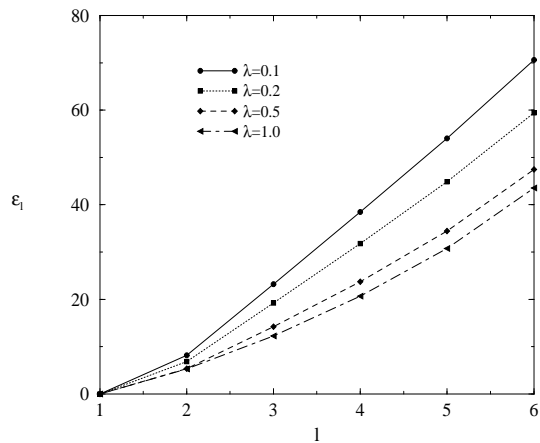}
\caption{Single-particle eigenvalues $\varepsilon_l$ 
 for the  first excited state of a transverse Ising chain
 for $L=12$ and four values of $\lambda$.
 } \label{fig5}
\end{figure}

  So far, we have treated the ground state, but one can also determine
  the density matrices for the first excited state $\mid \Phi_1>$. 
  This state contains an odd number of fermions. To apply the formalism
  here, one can perform a particle-hole transformation at one site, e.g.
  $c_1^{\dagger} \leftrightarrow c_1$. Then $\mid \Phi_1>$ appears in the
  even subspace and can be written in the form (\ref{eqn:gsconf}).
  With the help of the relations
  \begin{eqnarray}
    \eta_1^{\dagger} \mid \Phi_1> &=& 0\nonumber \\
    \eta_k \mid \Phi_1> &=& 0 \hspace{0.5cm} \mbox{for $k\geq 2$}
  \end{eqnarray}
  one can then derive the corresponding equation for the matrix $G_{ij}$.
  In this way, the single-particle eigenvalues $\varepsilon_l$  
  shown in figure \ref{fig5} were obtained. In contrast to the case of
  the ground state, the first eigenvalue is zero here. This reflects
  the fact that, in the original representation, the fermion number is odd,
  while the number of sites is even. The other eigenvalues
  are very similar to those for the ground state. In particular, one
  has a linear spectrum away from $\lambda = 1$ and a curved one at
  the critical point. The vanishing $\varepsilon_1$ 
  causes all eigenvalues $w_n$ of $\rho_1$ to be at least doubly
  degenerate.

 Finally, the closely related problem of the row-to-row transfer matrices 
for the two-dimensional Ising model can be studied in the same way. For a
square lattice with couplings $K_1$ ($K_2$) in the vertical (horizontal)
direction one can consider two symmetrized versions, namely
\be
V = {V_2}^{1/2} V_1 {V_2}^{1/2} ; \hspace{5mm}
      W = {V_1}^{1/2} V_2 {V_1}^{1/2} 
\ee
where $V_1$ ($V_2$) contain the vertical (horizontal) bonds. 
Both represent fermionic quantum chains and can be diagonalized 
also for open boundaries \cite{Abraham71,Kaiser/Peschel89}. For
the thermodynamics, one needs the eigenvector with maximal eigenvalue.
DMRG calculations using the operator $V$ have already been done \cite{Ising}.
The spectrum of the $\varepsilon_l$ in the isotropic case $K_1 = K_2$ is very 
similar 
to that found above in figure \ref{fig1}.
 This also holds for the magnitude of the 
$\varepsilon_l$ and the problem can therefore be treated equally well by DMRG. 
For $W$ the $\varepsilon$-spectrum is strictly linear at the lower end and
described by a formula containing elliptic integrals as in 
\cite{Pescheletal99}, while for $V$ the values are somewhat smaller and there
is a deviation from linearity for the first $\varepsilon_l$. This reflects
the difference in the representation of $\rho_1$ via CTM's in the two cases.

\section{XY-spin chain}
\label{sec:four}

In this section we consider briefly the spin one-half quantum chain described
by the Hamiltonian
\be
H = -J/2 \sum_{i=1}^{L-1} [(1+\gamma)\sigma_i^{x}\sigma_{i+1}^{x} +
    (1-\gamma) \sigma_i^{y}\sigma_{i+1}^{y}+
    h(\sigma_i^{z}+ \sigma_{i+1}^{z})] 
\label{equ:HXY}
\ee
which reads in terms of fermions
\be 
H = -J \sum_{i=1}^{L-1} [(c_i^{\dagger}c_{i+1} + 
 \gamma c_i^{\dagger}c_{i+1}^{\dagger}
   + h.c.) + h (c_i^{\dagger}c_i+c_{i+1}^{\dagger}c_{i+1}-1)]
\ee
Although similar to the transverse Ising chain, this system has a special
feature. For 
\be 
\gamma^2 + h^2 = 1
\label{equ:disorder}
\ee
the ground state simplifies and also becomes two-fold degenerate. In the spin
language, one has two simple product states \cite{Truong90}. Moreover,
the behaviour of correlation functions changes from monotonic to oscillatory
\cite{Barouch71} and thus (\ref{equ:disorder}) represents a "disorder line"
\cite{Stephenson70}. On this line, $H$ describes also a stochastic 
reaction-diffusion
model \cite{Hinrichsetal95} equivalent to Glauber's kinetic spin model.

The appearance of a simple ground state can be 
observed in the density-matrix spectrum and has already been seen in DMRG
calculations for certain other models (see section 3.1 in \cite{DMRG}).
 For the $XY$-chain,
it can be investigated very well in the fermionic approach.

\begin{figure}
 \hspace*{4cm}
     \epsfxsize=70mm
     \epsfysize=60mm
     \epsffile{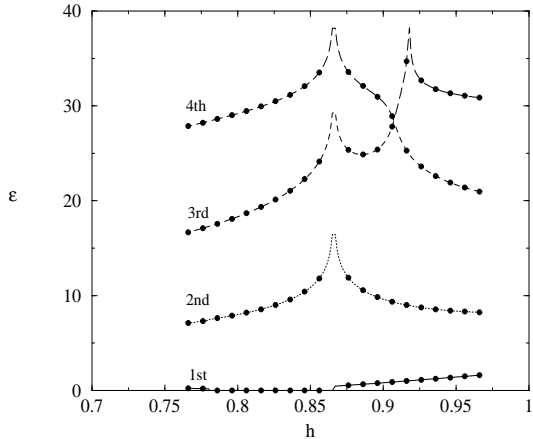}
\caption{The four lowest single-particle eigenvalues $\varepsilon$ 
   for an XY spin chain in a field $h$. The anisotropy is $\gamma=0.5$,
   the length $L=8$. Lines result from the analytical method, 
   solid circles from a DMRG calculation.}
   \label{fig6}
\end{figure}

In figure \ref{fig6} we show the lowest $\varepsilon_l$-values as a function
of the parameter $h$ for fixed $\gamma = 1/2$. The disorder point according to
(\ref{equ:disorder}) is then at $h_0 = 0.866$. One can see that, coming from 
larger 
values of $h$, all $\varepsilon_l$  except the lowest one diverge as one 
approaches
$h_0$. For $h < h_0$ they become finite again. In this region, 
however, one has to 
work in another subspace since at $h_0$ the lowest fermionic eigenvalue 
$\Lambda_0$
in $H$ crosses zero, which leads to the degeneracy of the ground state. 
This can
be done as for the excited state in Section \ref{sec:three}. 
Then one finds the curves in
the figure. As a check we also performed direct DMRG calculations and found 
complete
agreement (dots). Such crossings appear repeatedly as one reduces $h$ further.
The next one (for the chosen $L$) takes place at $h = 0.78$.
However, as seen from the figure, the higher $\varepsilon_l$ show no effects at
this point, indicating that the ground state of $H$ does not simplify there.
At $h_0$, the divergence of the $\varepsilon_l$ 
for $l \geq 2$ together with the
value $\varepsilon_1 = 0$ lead to the density-matrix eigenvalues $w_1 = w_2 = 
1/2$,
while all other $w_n$ are zero, i.e. 
the spectrum collapses at this point. This effect
could be a tool in the search for simple ground states by DMRG.

\section{two-dimensional tight-binding model}
\label{sec:five}

 As the last, but most important example we consider a
 tight-binding model with open boundaries described by
 \be
  H=-\sum_{<{\bf i,j}>} (c_{\bf i}^{\dagger} c_{\bf j} +
    c_{\bf j}^{\dagger} c_{\bf i})
  \label{eqn:tB}
 \ee
 where the brackets $<{\bf i,j}>$ denote nearest-neighbour sites. This model is
 critical and solvable in all dimensions. We treat it here for the case of a 
square 
 lattice and we assume that the system also has the shape of a square with
 $L = N^2$ sites where $N$ is even. This problem has has served as a DMRG test
 case some time ago \cite{Liang/Pang94}.

 The ground state here is different from that in the previous sections.
 Because $H$ only contains hopping, $B = 0$ in Equ.(\ref{eqn:Ham}),
  the fermion 
 number is fixed and $\mid \Phi_0>$ does not have the form (\ref{eqn:gsconf}).
 However, one can perform a particle-hole transformation on $L/2$ sites, 
 for example on every second one, by which the Hamiltonian acquires
 pair creation and annihilation terms ($B \neq 0$). Then $\mid\Phi_0>$, which
 originally contains $L/2$ particles, becomes a superposition of terms
 with particle numbers ranging from $0$ to $L$ and can again be written 
 in the form (\ref{eqn:gsconf}). In the same way, an arbitrary 
 $n$-particle eigenstate 
 of H could be handled by exchanging particles and holes at $n$ sites. 
 The density-matrix spectrum is not affected by such local 
 transformations.

 To carry out the calculation, one makes the problem formally one-dimensional
 by numbering the sites from 1 to $L$ in such a way that the desired partition 
 into
 two parts arises naturally. For example, a meander-like numbering as in 
 \cite{Liang/Pang94} permits a division of the square into two halves.

\begin{figure}
 \hspace*{4cm}
     \epsfxsize=70mm
     \epsfysize=60mm
     \epsffile{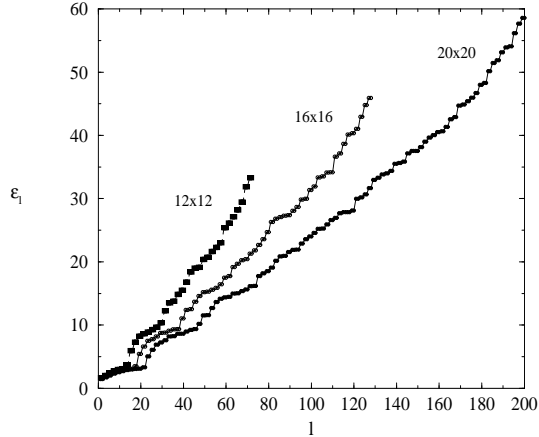}
\caption{Single-particle eigenvalues $\varepsilon_l$  for  
  two-dimensional tight-binding models of different sizes.  
  The $\varepsilon_l$ are for one half of the system. } \label{fig7}
\end{figure}

\begin{figure}
 \hspace*{4cm}
     \epsfxsize=70mm
     \epsfysize=60mm
     \epsffile{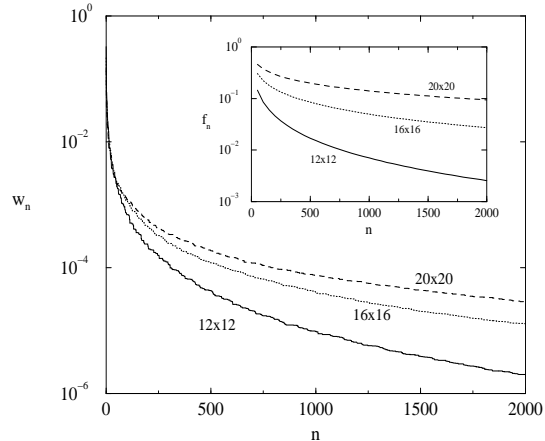}
\caption{Density-matrix eigenvalues $w_n$ of two-dimensional 
 tight-binding models, obtained from the $\varepsilon_l$ in Fig. \ref{fig7}.
 The inset shows the truncation error (see text).} 
 \label{fig8}
\end{figure}

 In figure \ref{fig7}, the single-particle eigenvalues $\varepsilon_l$ for such
 a half-system and three different sizes are shown. One notices two
 features which are in contrast to the one-dimensional results:
 a "foot" of low-lying $\varepsilon_l$ and a much smaller
 slope of the curves (note the scales). Both are strongly size-dependent.
 The number of $\varepsilon_l$ in the foot is equal to $N$, which
 indicates that these states are closely connected with the 
 interface between system and environment. 
 Figure \ref{fig8} shows the first 2000 eigenvalues $w_n$ which result.
 Due to the small $\varepsilon_l$, they decrease very slowly and 
 the situation worsens as the system is enlarged. The tails of the curves
 can be described qualitatively by $\ln(w_n) \sim -\ln^2(n)$ as in
 \cite{Okunishi99,Chung/Peschel00}. The effect of these tails shows
 up even more in the truncation error $f_n$, which is defined as the sum of
 all $w$'s beyond $n$. This quantity is given in the inset of the figure.
 With $n=2000$ it is approximately
 $5\times10^{-2}$, $5\times10^{-1}$ and $10^{-1}$, respectively. Thus the
 situation is not only much worse than for one-dimensional systems, but also 
 worse than for the two-dimensional system with a gap discussed in 
 \cite{Chung/Peschel00}. Standard DMRG calculations using, say, 2000 states
 would be limited to sizes below $12\times12$, and even then
 the accuracy would be much less than one is used to in quantum chains.

\begin{figure}
 \hspace*{4cm}
     \epsfxsize=70mm
     \epsfysize=60mm
     \epsffile{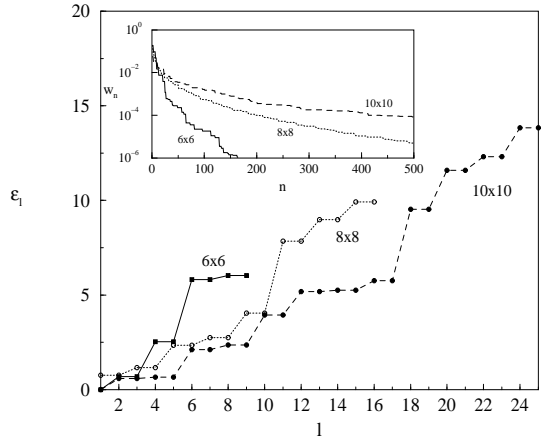}
\caption{Single-particle eigenvalues $\varepsilon_l$  of 
  two-dimensional tight-binding models. The $\varepsilon_l$ are for a quarter
  of the system, the $w_n$ obtained from them are plotted in the inset.} 
  \label{fig9}
\end{figure}

 One can also calculate the density-matrix spectra for other shapes of the
 selected subsystem. As an illustration, we show in figure \ref{fig9}
  results for one 
 quarter of a quadratic system (for example the upper right one). Note that
 the sizes indicated there refer to the whole system. One sees again some
 small eigenvalues, but fewer than for the half-system, while there are
 further higher-lying plateaus and additional short steps. Obviously this 
reflects
 the particular interface with a corner. For the $10\times10$ system, for 
 example, the two lowest plateaus contain 9 states which is just the number
 of sites along the interface. The eigenvalues $w_n$ are plotted in the
 inset of the figure. They are similar to those for the half-system but some 
 more steps persist for small $n$.
 In the same way, one can investigate cases where one cuts the
 square diagonally at various positions. 
 Such partitions appear in a recent new 
 DMRG algorithm \cite{Xiang01}. 
 The general features of the spectra, however, do not
 change.

 Finally, let us mention that one can also include spin in $H$
 and thereby treat the Hubbard model in the $U=0$ limit. 
 Then the operators $f_l$, $f_l^{\dagger}$ in $\rho_1$ 
 acquire a spin label, too,
 and all single-particle levels become doubly degenerate. 
 This makes the tails of the $w_n$-curves even flatter than in the 
 spinless case. However, the curves are also pulled down by smaller
 normalization factors which leads to a faster initial decay.
 For a $20\times20$ lattice, the spectrum of the first $3000$ states is,
 on the whole, rather close to that shown in figure \ref{fig8}.

\section{Conclusion}
\label{sec:six}

We have studied the reduced density matrices for non-interacting
fermions on a lattice. The key ingredient for the calculation was a
simple representation of the (ground) state. This led rather directly
to the exponential Boltzmann-like form of the density matrices. The only
 really numerical step involved was the calculation of the single-fermion
eigenvalues appearing in the exponent. With these, we discussed a number
of cases in one and two dimensions with characteristic differences.
We focused on the eigenvalues, but one can also investigate the single-
fermion eigenfunctions. One then sees that they are concentrated near the
interface between the two parts of the system. This explains the decisive
role of the connectivity for the spectra.

One should mention that fermionic density matrices have been studied before, 
e.g. 
in quantum chemistry \cite{Coleman63,Davidson76}. However, in this case the systems
are continuous and the Hilbert space is infinite. Then already the single-particle 
density matrices have inifinitely many eigenstates \cite{Davidson62}.
Our systems are discrete, but we are interested in density matrices for 
arbitrarily
large subsystems. These are non-trivial even for non-interacting fermions.
From the experience with other models, one can expect that the results are 
roughly
representative also for more complicated systems.

For this reason, the two-dimensional case is particularly important.
With our formulae, we could treat the tight-binding model for arbitrary 
partitions
of the system. This allows to make much more detailed statements than a 
previous, purely numerical investigation of this system \cite{Liang/Pang94}.
In particular, one can see the very slow decay of the spectra and of
the truncation errors directly. Basically, it is connected with the existence
of long boundaries between the two parts of the system. In the current
DMRG procedures, these appear necessarily at some point of the calculation.
Therefore it is not yet clear whether a recent new algorithm \cite{Xiang01}
can really overcome this problem.

\section*{Acknowledgement}
We are indebted to H. Grabert who drew our attention to fermionic coherent 
states.
M.C.C. thanks C.T.Chan for discussions and 
the Deutscher Akademischer Austauschdienst (DAAD) for financial
support.

  \section*{Appendix}
   Here we list some details concerning the steps 
   in section (\ref{sec:two})   

   (A) To derive (\ref{eqn:gsrelation}), one writes the Eqn. 
   (\ref{eqn:eta}) explicitly as
  \begin{equation}
   \sum_{n} (g_{kn} c_n + h_{kn} c_n^{\dagger}) e^{F} \mid 0> = 0
   \label{eqn:gsrel}
  \end{equation} 
   where $ F= 1/2 \sum_{ij} G_{ij} c_i^{\dagger} c_j^{\dagger}$. Using the 
   relation
  \be
   [c_i,e^{F}]=\frac{\partial}{\partial{c_i^{\dagger}}} e^{F}
  \ee
  the exponential factor can be brought to the left
  \be
   e^{F} \sum_{n} \{\sum_{m} g_{km} G_{mn} + h_{kn}\} c_n^{\dagger} 
   \mid 0> =0 \hspace{0.5cm}  \label{eqn:gsrel2}  
   \ee
   Since this must  hold for all $k$, 
   the only possibility is that the term in the bracket vanishes
   which gives the desired result.

   (B) The explicit form of the integrand 
    in (\ref{eqn:integral})
   is
   \be
   \exp{\{-{\xi_2^{\ast}}^T\xi_2 + 1/2({\xi_2^{\ast}}^Ta^{22}\xi_2^{\ast} 
    -{\xi_2}^Ta^{22}\xi_2)-({\xi_1^{\ast}}^Ta^{12}\xi_2^{\ast}+
    {\xi_2}^Ta^{21}\xi'_1)+1/2({\xi_1^{\ast}}^Ta^{11}\xi_1^{\ast}-
     {\xi'_1}^Ta^{11}{\xi'_1})\}} \label{eqn:integrand}
   \ee 
   where $\xi_1^{\ast}, \xi'_1 (\xi_2^{\ast}, \xi_2)$
   are vectors composed of the variables of  part 1 (part 2), respectively.
   Using the notation $\xi\equiv (\xi_2,\xi_2^{\ast})$, 
   this can be rewritten as 
   \be
    \exp{\{-\xi^{\dagger}\hat{B}\xi + 
    \zeta^{\dagger} \xi + \xi^{\dagger}
    \eta +\hat{K}}\} \label{eqn:gauss}
   \ee  
   where $\hat{B}$ is a $2(L-M)\times 2(L-M)$ matrix containing $a^{22}$, 
   $\zeta,\eta$ are both $2(L-M)$ dimensional vectors constructed from
    $a^{12},a^{21},
   \xi_1^{\ast}$ and $\xi'_1$ and $\hat{K}$ is the last term 
   in (\ref{eqn:integrand}). 
    (\ref{eqn:gauss}) is an explicit 
   Gaussian form which can be integrated whereby  
   (\ref{eqn:gve}) is obtained.

   (C) To derive the operator form for $\rho_1$ from Eqn. (\ref{eqn:gve}),
   one first diagonalizes the matrix $\beta$.
   This transforms (\ref{eqn:gve}) into a similar form with modified matrix 
   $\alpha$.
    Using the relations 
   \begin{eqnarray} 
   <\xi_i\xi_j\mid c_i^{\dagger}c_j^{\dagger} 
   &=& <\xi_i\xi_j\mid \xi_i^{\ast} \xi_j^{\ast}\nonumber\\
    c_ic_j\mid\xi'_i,\xi'_j> &=& 
   \xi'_i\xi'_j\mid\xi'_i\xi'_j>
   \end{eqnarray}      
   one can replace $\xi_i^{\ast}\xi_j^{\ast}$ with 
   $c_i^{\dagger}c_j^{\dagger}$ and $\xi'_i\xi'_j$
   with $c_ic_j$ in the left and right exponentials. The cross terms 
   $e^{\lambda_i\xi_i^{\ast}\xi'_i}$, where $\lambda_i$ is one of the 
   eigenvalues of 
   $\beta$, can be rewritten with the relation
   \be
     <\xi_i\mid f(c_i^{\dagger},c_i) \mid\xi'_i> = e^{\xi_i^{\ast}\xi'_i}
    f(\xi_i^{\ast},\xi'_i)\label{eqn:replace}
   \ee 
    In our case  the left-hand side equals 
    $e^{\lambda_i \xi_i^{\ast}\xi'_i}=1+\lambda_i\xi_i^{\ast}\xi'_i$
     so that
   \begin{eqnarray}
    f(c_i^{\dagger},c_i) &=& (1+(\lambda_i-1)c_i^{\dagger}c_i)\nonumber\\
    &=& e^{\ln{\lambda_i} c_i^{\dagger}c_i}
   \end{eqnarray} 
    Transforming  back to the original representation leads to
    (\ref{eqn:rho1o}).  
   
    (D) The operator $\rho_1$ in (\ref{eqn:rho1o})
     can be diagonalized by calculating the Heisenberg operators
    $\rho_1c_j\rho_1^{-1}$ and $\rho_1c_j^{\dagger}\rho_1^{-1}$
    as in \cite{Kaiser/Peschel89}.
    Due to the form of $\rho_1$, they are linear combinations of the $c$
    and $c^{\dagger}$. Inserting the Bogoliubov transformation 
    and following \cite{Kaiser/Peschel89} one finds that the 
    eigenvalues $\varepsilon_l$ can be obtained from the equation 
    \be
     (\beta+\beta^{-1}+\beta^{-1}\alpha-\alpha\beta^{-1}
     -\alpha\beta^{-1}\alpha)\; \chi_l = 2\cosh{\varepsilon_l}\; \chi_l
    \label{eqn:eigen}
    \ee    
    Typically, the matix $\beta$ has elements which vary exponentially
    over a large range. This limits the size of the systems for which one
    can use Eqn. (\ref{eqn:eigen}) in actual numerical calculations.

\end{document}